\documentclass[11pt]{article}
\usepackage[utf8]{inputenc}
\usepackage{amsmath}

\usepackage{algcompatible}
\usepackage{graphicx}
\usepackage[ruled]{algorithm2e}
\usepackage{soul}
\usepackage{amsfonts} 
\usepackage{comment}
\usepackage{todonotes}
\usepackage{cases}
\usepackage{appendix}
\usepackage{multirow}
\usepackage{adjustbox}
\usepackage{multicol}
\usepackage{booktabs}
\usepackage{float}
\usepackage{enumitem}

\newtheorem{definition}{Definition}

\definecolor{CF5F5F5}{HTML}{F5F5F5}
\definecolor{C000000}{HTML}{000000}
\definecolor{CEBEBEB}{HTML}{EBEBEB}
\definecolor{C0000FF}{HTML}{0000FF}
\definecolor{CFFCC00}{HTML}{FFCC00}
\definecolor{CFF0000}{HTML}{FF0000}
\definecolor{CFFFF00}{HTML}{FFFF00}
\def\nod [#1]{\node[#1] {} }

\title{A Multi-Objective Degree-Based Network Anonymization Approach}
\author{Ola N. Halawi ~and Faisal N. Abu-Khzam\\
\\
Department of Computer Science and Mathematics\\ 
Lebanese American University\\ 
Beirut, Lebanon}
\date{}
\begin{document}
\maketitle

\thispagestyle{empty}

\begin{abstract} 
Enormous amounts of data collected from social networks or other online platforms are being published for the sake of statistics, marketing, and research, among other objectives. The consequent privacy and data security concerns have motivated the work on degree-based data anonymization. 
We propose and study a new multi-objective anonymization approach that 
generalizes the known degree anonymization problem and attempts at improving 
it as a more realistic model for data security/privacy. 
Our suggested model guarantees a convenient privacy level based on modifying the degrees in a way that respects some given local restrictions, per node, such that the total modifications at the global level (in the whole graph/network) are bounded by some given value. The corresponding multi-objective graph realization approach is solved using Integer Linear Programming to obtain the best possible solutions. Our experimental studies provide empirical evidence of the effectiveness of the new approach; by specifically showing that the introduced anonymization algorithm has a negligible effect on the way nodes are clustered, thereby preserving valuable network information while significantly improving data privacy.
\end{abstract}

%\begin{keyword}
%Data privacy,
%Network data security,
%Anonymization,
%Degree-based anonymization 
%\end{keyword}

\section{Introduction}

Among the available privacy-preserving techniques for network-based data, degree anonymization proved to be a practical tool in terms of conserving data utility and resisting re-identification attacks. It works by altering the set of edges of a graph so that nodes are indistinguishable in terms of their degrees. Formally, a graph $G$ is $k$-degree anonymous if for every vertex $v$ there are at least $k-1$ other vertices with the same degree as $v$ \cite{kneighborhood}. This particular type of ``hiding in the crowd'' guarantees privacy because an attacker can identify an individual with a probability of less than $\frac{1}{k}$, where $k$ is the anonymity\slash security level desired by the data publisher.

Unfortunately, the above classical definition of degree anonymity is too restrictive. Requiring at least $k$ vertices to have the same degree, for each possible vertex-degree, could have a very high edge-modification cost in large networks.
An alternative multi-objective optimization approach is proposed in this paper. Our model sets restrictions on the number of added and deleted edges per-vertex and relaxes the same-degree restriction by setting a range parameter on the resulting degrees so they would be required to be close enough, while not affecting the privacy level. Thus, it extends the real applicability of this type of network data anonymization. We formally define degree anonymization as a multi-objective optimization (or multi-parameterized) problem as follows.

\vspace{10pt}
\noindent
{\bf Multi-Parameterized Degree Anonymization}

\vspace{5pt}
\noindent
{\bf Given:} An undirected graph $G=(V,E)$, positive constraint-parameters $a,d$, a range-parameter $t$ and an anonymization-parameter $k$.

\noindent
{\bf Question:} Can we obtain an anonymous graph $G'=(V,E')$ by adding at most $a$ and deleting at most $d$ edges per single vertex, so that for each vertex $v$ we have at least $k-1$ other vertices whose degrees fall into the interval $[degree(v)-t,degree(v)+t]$?

\vspace{5pt}

The anonymization parameter $k$ is assumed to be pre-defined by the user based on the desired anonymization level and the type of the given data. 
The parameters $a$, $d$, and $t$ are computed by the algorithm, as we propose in this paper, to deliver the least cost target solution.
It is worth noting that adding local parameters or constraints, typically in graph modification problems, is known to have a notable positive effect on the problem's complexity while improving the practicality of the model. This was noted recently in \cite{multi-param} in the context of Cluster Editing, which inspired the above formulation and the work we present in this paper.

The above-proposed model is a generalization of the $k$-degree anonymization problem. The resulting graph realization problem can be viewed as a new weighted variant of the {\em Edge Cover} problem, which we formulate as an Integer Linear Programming problem. We test our approach by applying clustering on the initial graph and the modified one after performing our anonymization procedure. The comparison is based on measuring the symmetric difference between clusters. Besides, we compare some graph metrics before and after perturbation.

\section{Preliminaries}

We assume familiarity with basic graph theory terminology such as adjacency, vertex degree, and neighborhood, among others. Networks, or graphs, are often subject to attacks that are based upon some prior or presumed knowledge about the degrees of some targeted nodes. This is assumed to be one of the most significant types of attacks \cite{arvind} since it could threaten individuals' privacy by inferring knowledge about the links to a node and the topology of the graph. Other types of attacks that rely on a priori knowledge of the neighborhood of a targeted node, or the corresponding induced subgraph, can be reduced down to knowledge about the degree of that targeted node.

Each graph $G$ can be represented by a sorted (often decreasing) sequence of the degrees of its vertices. We refer to this sequence by $D(G)$, being unique for $G$.
The notion of the $k$-degree anonymity of a graph can thus be reduced to a transformation of its degree sequence into an anonymized one. We seek a minimal number of edge-editing operations.

\begin{definition}
Generally, a degree sequence $D(G)$ is $k$-anonymous if every value in $D(G)$ is repeated at least $k-1$ times. Then, a graph $G$ is $k$-degree anonymous if its degree sequence is $k$-anonymous.
\end{definition}

Our approach is divided into two parts. In the first, the descending degree sequence $D(G)$ is computed and anonymized to produce a degree sequence $D'$. In the second part, $G'$ is constructed so that $D' = D(G')$.
This latter procedure is known as graph realization, which we formulate as an ILP problem. 

The first degree-based anonymization approach seems to be due to Liu and Terzi \cite{liu} who proposed a technique that modifies a given graph/network by adding\slash deleting a certain number of edges to generate a $k$-degree anonymous graph. 
While the work of Liu and Terzi is motivated by logical intuitions, they admit that additional work needs to be done to develop theoretically and practically sound privacy models for graphs.

In some other attempts, the set of vertices is changed instead of the set of edges. For example, Chester et al. proposed an approach that generates new edges between auxiliary and real nodes or between auxiliary nodes \cite{chester}. For unlabeled graphs, experimental results demonstrated that perturbing the set of vertices changes some important properties of a graph (such as how the nodes cluster) and weakens the ``data accuracy'' because of the information loss.

In \cite{univariate}, Casas-Roma et al. presented an algorithm for graph anonymization based on the univariate micro-aggregation. It works by modifying the set of edges based on the univariate micro-aggregation method for data protection. Another (different) approach was recently presented in \cite{dcc-poster} based on a graph compression scheme. It consists of mapping large networks into points in a certain high-dimensional space using the (distributed) FastMap algorithm (see \cite{abu02, faloutsos}). 

A more recent approach of Alavi et al. named the ``GAGA Graph Anonymizer,'' appeared in 2019 \cite{Gaga}. Their ``Genetic Algorithm for Graph Anonymization'' is claimed to be the best solution for networks’ protection because it overcomes all the limitations of other available solutions. Another genetic algorithm was presented in \cite{jordiGA}. Genetic algorithms, however, are known for their limitation in guaranteeing an optimal (or near-optimal) solution. Besides, with the increase of problem size, especially with real large networks, a genetic algorithm tends to slowly converge to a local minimum.
In our work, we can achieve optimum solutions, modulo the various parameters, by employing an ILP formulation of the problem.

\section{Generating $k$-Anonymous Degree Sequences}

We present an algorithm that takes an original graph $G$ and privacy level $k$ as input and starts by generating the descending ordered degree sequence $D(G)$ and copying it into the target degree sequence $D'$ where the degree-modifications are performed. 
We assume that the privacy level $k$ is given. Otherwise, we compute a privacy level that is most suitable for $G$ based on the deviation of the degrees in $D$. We leave a detailed description of this latter approach for future work.
We apply a vector-based approach that divides $D'$ into chunks of at least $k$ nodes. For every segment, we define the constraint parameters $a$ and $d$, which correspond to the maximum number of edge additions and deletions per single vertex, respectively. We set $a$ as the absolute value of the difference between the first degree in each segment and the last one minus $2*t$ and we set $d = a$ as a default value. A user of our algorithm/code might prefer to distinguish between the two values (in which case $d$ can be adjusted at this stage).

The average degree in each part is calculated, and the node's degree that is the closest to the average will be set as the degree of the first node in that chunk in $D'$. 

\begin{algorithm} [hbt!]
\SetAlgoLined
\textbf{Input:} $G=(V,E), k$ ;\\ 
Generate $D(G);D'\longleftarrow D(G)$;\\
$n \longleftarrow |V|;  add \longleftarrow 1;  delete \longleftarrow 1$;\\
\For {$i = 1; i < n; i=i+k$} {
$a=|D'[i]-D'[i+k+1]-2*t|$;\\
$d=a$;\\
$avg \longleftarrow \text{Average of degrees}$;\\
$j \longleftarrow \text{Index of the closest node to avg}$;\\
$D'[i]=D'[j]$;\\
$lowerbd \longleftarrow D'[i]-t$;\\
$upperbd \longleftarrow D'[i]+t$;

\ForEach{$z \in [i+1, i+k]$}{
\If{$D'[z] \notin [lowbd, upperbd]$}{
$changes \longleftarrow |lowerbd-D'[z+1]|$\\
\If{$changes \leq a$}{
\While{$add \leq a$ \text{and} $add \leq changes$}{
$D'[z+1]=D'[z+1]+add$;\\
$add++$;
}}
\If{$changes > a$}{
\While{$delete \leq d$; $add \leq a$ \text{and} $add \leq changes$ \textsc{and} $D'[z] \geq D'[z+1]$}{
$D'[z]=D'[z]-delete$;\\
$D'[z+1]=D'[z+1]+add$;\\
$add++$;\\
$delete++$;
}}}
{$D'[z+1]=D'[z]$};
}
}
return $D'$
\caption{\textbf{Multi-parameterized $k$-degree anonymization}}
\end{algorithm}

If the degree of the next node does not fall into the interval $[degree(v_1 )-t, degree(v_1 )+t]$, then we add for that node the required number of edges for its degree to belong to the above interval. If the number of edges that have to be added exceeds the limit $a$, then we delete edges from the previous node and add edges to the second node within the allowed range. The algorithm will repeat the same process until $D'$ is all covered.

%\noindent
%\underline{\bf Running Time and Complexity:} 
From a time-complexity standpoint, the worst-case scenario happens when there is a high standard deviation between the degrees of nodes, or when all nodes have different degrees. In this case, many edges will be affected to reach the anonymity level of $k$. For each node, we may have at most $k-t$ operations of edge addition and/or deletion. Therefore, the time complexity is $O(k*n)$. In the best case, on the other hand, there would be no significant divergence between the degrees of nodes, or the graph is nearly regular (all nodes have almost the same degree), then the time complexity would be $O(n)$.

\section{A Graph Realization Approach}

In the second phase of the anonymization procedure, the output graph $G'$ is constructed based on the target degree sequence $D'$. Recall that we aim at producing a graph that is $k$-anonymous by performing the least amount of perturbation to the topology and structure of $G$. The common practice is to remove edges incident on vertices that have to decrease their degrees and add new edges to vertices that have to raise their degrees, without taking into consideration the major effect of graph re-construction on the topology of the network. We, on the other hand, define ``the realization'' as an optimization problem formally as follows.

\vspace{10pt}

\noindent
{\bf Weighted Graph Realization}

\noindent
\underline{\bf Given:} A graph $G$ and a function $r:V(G) \rightarrow \mathbb{Z}$.

\noindent
\underline{\bf Question:} Is there a sequence of edge editing operations that results in $G' = (V,E')$ such that $\forall v \in V(G'), degree_{G'}(v)=degree_G(v)+r(v)$?

\vspace{5pt}

The realization procedure we propose begins by copying $G$ into $G'$ on which edge edits are performed. Then, it computes the vector of changes $\theta$ between $D'$ and $D$, where $\theta=D'-D$. This vector is used to detect which nodes have to introduce or remove incident edges. It can also be used to compute the anonymization cost. To further explicate, if $\theta[v]$ is negative, then $|\theta[v]|$ incident edges on $v$ should be removed. However, if $\theta[v]$ is positive, then $\theta[v]$ edges should be added to $v$. 
Each vertex will be labeled by its corresponding weight in $\theta$.

This graph realization problem can be seen as a vertex-weighted variant of the Edge Cover problem (VWEC). Classically, a set of edges $C \subseteq E$ is an edge cover of $G$ if every vertex in $V$ is incident on at least one edge in $C$. VWEC can be defined formally as follows.

\vspace{10pt}
\noindent{\bf Vertex-Weighted Edge Cover}

\noindent
\underline{\bf Given:} A graph $G=(V,E)$ and $w:V \rightarrow \mathbb{Z}$;
%where $w(v)$ is the weight of vertex $v$.

\noindent
\underline{\bf Question:} Is there a set of edges $C$ in $E$ such that every vertex $v \in V$ is incident on $w(v)+deg_{G}(v)$ edges in $C$?   

\vspace{5pt}

It is well known that {\sc Edge Cover} is solvable in polynomial-time. However, VWEC is NP-Complete by a trivial reduction from {\em Editing to a Given Degree Sequence}, which is shown to be NP-hard in \cite{GOLOVACH20171}. We formulate this particular graph realization problem as an Integer Linear Programming problem (ILP). For this purpose we use a variable $x_{i,j}$ for every pair of vertices $i$ and $j$, such that $x_{i,j}\in \{0,1\}$.
%taking the values $0$ and $1$. 
The interpretation is that $x_{i,j}=1$ if $\{i,j\} \in C$, otherwise $x_{i,j}=0$.
%ot means that $\{i,j\} \in C$. 
%The cost of the solution, which we want to minimize, is $\sum\limits_{i\in V, j\in V} {x_{ij}}$. 
This gives the following ILP formulation:

\vspace{10pt}

\begin{equation*}
\begin{aligned}
&\text{Minimize}
\sum\limits_{i\in V, j\in V} {x_{ij}}\\ 
&\text{Subject to:}\\
&\sum\limits_{j\in V} {x_{ij}}= |\theta[i]| ~ ~\forall i\in V
\end{aligned}
\end{equation*}

%Minimize $\sum\limits_{i\in V, j\in V} {x_{ij}} %\text{ such that:}
%\begin{gather*}
%\begin{cases}
%\{i,j\}\notin E \text{ if } \theta[i]>0\\
%\{i,j\}\in E \text{ if } \theta[i]<0\\
%\end{cases}
%\text{Subject to: } $\sum\limits_{j\in V} {x_{ij}}= |\theta[i]| ~ ~\forall i\in V$
%\end{gather*}

\begin{comment}
\noindent
If $\theta[i]>0$ :
\begin{gather*}
\begin{cases}
Minimize \sum\limits_{i\in V, j\in V} {x_{ij}} \text{ such that} \{i,j\}\notin E\\
\text{Subject to } \sum\limits_{i\in V, j\in V} {x_{ij}}= \theta[i]
\end{cases}
\end{gather*}
\\

\noindent
If  $\theta[i]<0$ : 
\begin{gather*}
\begin{cases}
Minimize \sum\limits_{i\in V, j\in V} {x_{ij}} \text{ such that} \{i,j\}\in E\\
\text{Subject to } \sum\limits_{i\in V, j\in V} {x_{ij}}= |\theta[i]|\\
\end{cases}
\end{gather*}
\end{comment}

\vspace{5pt}

%The realization procedure distinguished between the  makes sure the total number of added edges sum of the weight by considering the vertices with positive 

%The realization process depends on the sign of the weight-function $\theta$ at each vertex. For the vertices with positive $\theta$-values, we need to add new incident edges to each of them. For the vertices with negative $\theta$-values, we remove edges incident on each. Overall the obvertex $i$ the function will minimize the sum of the weights of edges that are incident on $i$ such that $\sum {x_{ij}}= |\theta[i]|$ where $j$ is a neighbor of $i$. 
%That is to say, we search for the weighted maximum minimal edge cover using the non-adjacent nodes of $i$.

%To achieve this, the objective function minimizes the sum of the weights of the ``non-edges'' such vertices.
%such that $\sum {x_{ij}}= \theta[i]$ where $j$ is not adjacent to $i$. 

%create new edges to $i$, 
%the objective function minimizes the sum of the weights of the none-edges such that 

%In this case, the algorithm searches for the weighted maximum minimal edge cover using the incident edges of $i$. 
%We take into consideration the significance of edges to be deleted to the connectivity of the entire network. Hence, the least significant edges will be deleted while the critical ones will be protected.

The realization algorithm will output a graph $G'$ with a degree sequence $D'$ if it exists. To avoid having negative answers, that is cases where no solution can be found, we use our introduced relaxed version of the problem (as discussed before). In the modified ILP formulation it is assumed that most of the edges of $G$ appear in $G'$ but not exactly all of them. This will be done by adding a threshold $y_i$ to $\theta[i]$, $\forall i \in V$. The minimum $y_i$ is in fact the minimum $t$ so that $D'$ is realizable. Each vertex will be labeled by $\theta[i]+y_i$ such that $y_i$ is the smallest possible value that can make $D'$ realizable. We obtain the following relaxed ILP:

\begin{equation*}
\begin{aligned}
&\text{Minimize} \sum\limits_{i\in V, j\in V} {x_{ij}} + \sum\limits_{i \in V}{y_i}\\
&\text{Subject to:}\\ &\sum\limits_{i\in V, j\in V} {x_{ij}}= |\theta[i]+y_i|
\end{aligned}
\end{equation*}

\section{Experimental Analysis}

We implemented our multi-objective degree anonymization algorithm and ran multiple tests using different graphs to assess its efficiency and, most importantly, its ability in preserving data utility. It was implemented in $Java$, and we used $CPLEX$ for the realization algorithm.
The tests were performed on an Intel(R) Core(TM) i7-7600U CPU $@$ 2.80 GHz machine. 

We compared our results with the ones presented in \cite{ying} for the algorithm of Liu and Terzi ($k$-Degree Anonymization: $k-DA$). We also compared our algorithm to the vertex addition approach by Chester et al.\cite{chester} in which the anonymization process takes place by adding fake vertices to the network and fake edges between these vertices and between fake vertices and real ones to attain the required $k$-anonymity level. Finally, we compared our results with Jordi-Casas et al.'s results in \cite{univariate}. Their algorithm is based on the univariate micro aggregation. It is available in two versions. The first, dubbed ``NC'', is based on the notion of neighborhood centrality to decide which edges to delete. It tries to preserve the edges that are more relevant to the connectivity of the whole network than others. The second approach (``Rand'') randomly selects edges to be removed.

%one is named ``NC" and the second one is ``Rand". They both work on altering the set of edges, but NC relies on a measure called neighborhood centrality to decide on which edges to delete. It tries to preserve the edges that are more relevant to the connectivity of the whole network than others. On the other hand, Rand chooses edges to be removed randomly.  

Real data sets were used for testing. The selected networks were obtained from the Konect Project \cite{konect} and \cite{NetworkRepository}. They are different in terms of topological, structural, and attribute properties. However, all the graphs used are simple, unweighted, and undirected.  

The notion of data utility is not well defined yet, so different authors have different approaches in quantifying data loss in a graph. In our work, we use the following graph structural measures to quantify and analyze information loss induced by the anonymization process:

\begin{itemize}

\item The largest eigenvalue of the adjacency matrix implies information about the diameter of a network and its cycles.

\item The second smallest eigenvalue of the Laplacian matrix implies information about the tree structure of the graph. It shows if the communities separate efficiently or not.

\item The average distance is the average of the shortest paths between all nodes in a network.

\item Harmonic mean of the shortest distance is similar to the average distance, it is used to evaluate the connectivity of a network.

\item Modularity measures the strength of the division of a network into clusters. High modularity values imply dense connections between the nodes of a graph.   
    
\item Transitivity is similar to the clustering coefficient; it detects the presence of loops near a vertex.
     
\item Subgraph centrality measures the number of subgraphs a vertex takes part in, weighting them according to their size.

\end{itemize}

\subsection{Empirical Results}

\begin{figure}[H]
\centering
\includegraphics[width=\textwidth,height=\textheight,keepaspectratio]{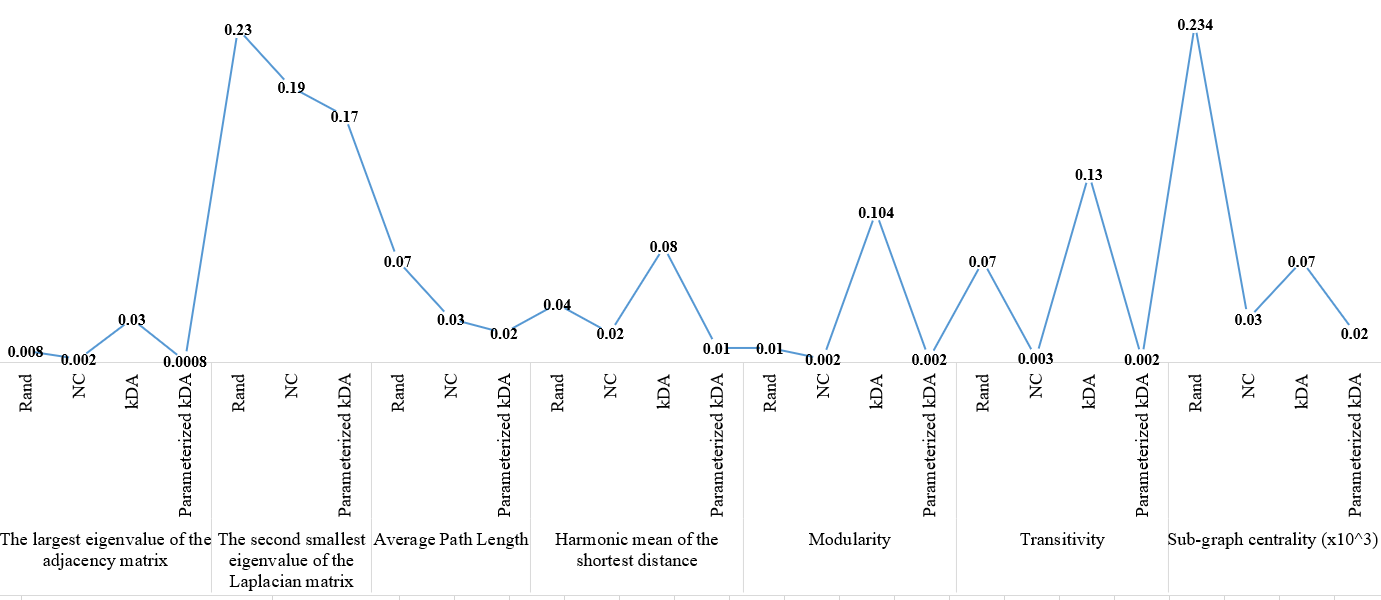}
\caption{Polbooks experimental results}
\label{fig:polbooks-results}
\end{figure}

We tested each dataset using the above-described 
algorithms in addition to our multi-parameterized model. We used the "Polbooks" network, it was not used to test the vertex addition algorithm, so we compared our multi-objective approach with Rand, NC, and kDA. We recorded the error induced by each algorithm on different utility measures. As shown in figure \ref{fig:polbooks-results}, Rand and NC algorithms surpass kDA, which induces a marginally less average error on the value of the transitivity measure. In general, NC outperforms Rand because it keeps essential edges in the network, which preserves the values of the measures to some extent. By comparing the average error produced by each algorithm, it is noticeable that our algorithm induces less noise and keeps the measures very close to their original values.

\begin{figure}[H]
\centering
\includegraphics[width=\textwidth,height=\textheight,keepaspectratio]{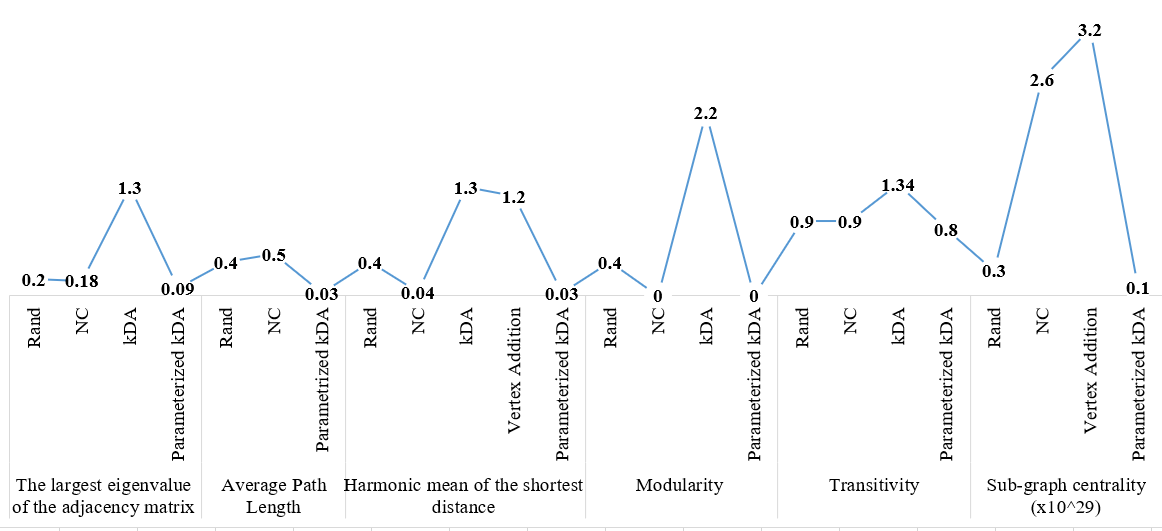}
\caption{Polblogs experimental results}
\label{fig:polblogs-results}
\end{figure}

We tested a larger network which is ``Polblogs.'' According to figure \ref{fig:polblogs-results}, Rand and NC produced an average error that is less than that produced by Vertex Addition or kDA on all measures. kDA produced a remarkable deviation from original measures values. The average error is 0.286 for NC and 0.291 for Rand while it is 1.953 for kDA on the largest eigenvalue of the adjacency matrix. Keeping eigenvalues close to their original values implies preserving cycles and the diameter of the network. Our approach outperformed Rand and NC by producing a smaller average error of around 0.213. Likewise, the vertex addition algorithm produces a 0.043 average error on the harmonic mean of the shortest distance while it is 0.0037 using our algorithm. Preserving the value of the harmonic mean of the shortest distance implies preserving connectivity and path lengths. The same analysis applies to other measures.

%\vspace{-20mm}

\subsection{Clustering Analysis}

To test the impact of the anonymization process on the topology of graphs and knowledge extraction, we measure its effect on how data elements cluster.
%used clustering. 
%The main rationale stems from the fact that preserving how data elements cluster is essential for preserving data utility.
%The objective is to check the effect of anonymization on how the data elements cluster. 
If the anonymization algorithm keeps almost the same clustered communities, then the topology of the original graph is not affected much and the released graph would be useful for data mining applications. To compare the clusters of the original graph $G$ and those of the anonymized graph $G'$, we need a certain measure of divergence. We used a precision index defined in \cite{index}, which has a value between 0 and 1. The closer to 1 the closer are the clusters of $G$ and $G'$. A value of 1 is obtained when the clusters of $G$ and $G'$ overlap completely.
The corresponding experiments were implemented using RStudio\cite{rstudio} and the results were recorded in Table \ref{tab:clustering}. 
The best algorithm is the one that has the highest precision index or the least precision error. 

%The clustering algorithm we applied is the Fast Greedy \cite{fastgreedy}, which detects communities based on modularity optimization. It starts with a sub-network of edges between highly connected vertices. Iteratively, it optimizes the modularity of the sub-network by sampling and adding random edges.

\begin{table}
  \centering
  \caption{The precision error produced by our approach versus that of NC on different data sets}
    \begin{tabular}{rrrrr}
          &       &       &       &  \\
\cmidrule{2-4}    \multicolumn{1}{r|}{} & \multicolumn{1}{l|}{Network } & \multicolumn{1}{l|}{Algorithm} & \multicolumn{1}{l|}{Precision Error} &  \\
\cmidrule{2-4}    \multicolumn{1}{r|}{} & \multicolumn{1}{l|}{\multirow{2}[2]{*}{American college football}} & \multicolumn{1}{l|}{NC} & \multicolumn{1}{r|}{0.053} &  \\
    \multicolumn{1}{r|}{} & \multicolumn{1}{l|}{} & \multicolumn{1}{l|}{Parameterized kDA} & \multicolumn{1}{r|}{0.003} &  \\
\cmidrule{2-4}    \multicolumn{1}{r|}{} & \multicolumn{1}{l|}{\multirow{2}[2]{*}{Erdos}} & \multicolumn{1}{l|}{NC} & \multicolumn{1}{r|}{0.187} &  \\
    \multicolumn{1}{r|}{} & \multicolumn{1}{l|}{} & \multicolumn{1}{l|}{Parameterized kDA} & \multicolumn{1}{r|}{0.114} &  \\
\cmidrule{2-4}    \multicolumn{1}{r|}{} & \multicolumn{1}{l|}{\multirow{2}[2]{*}{Enron}} & \multicolumn{1}{l|}{NC} & \multicolumn{1}{r|}{0.121} &  \\
    \multicolumn{1}{r|}{} & \multicolumn{1}{l|}{} & \multicolumn{1}{l|}{Parameterized kDA} & \multicolumn{1}{r|}{0.081} &  \\
\cmidrule{2-4}          &       &       &       &  \\
    \end{tabular}
  \label{tab:clustering}
\end{table}

%Clustering is very essential in analyzing graphs' structures and data mining information. 

According to \cite{univariate}, NC produces less precision error than Rand on most data sets. However, our multi-parameterized degree anonymization algorithm surpasses NC by notably minimizing the precision error. For the American college football network, our precision error is $43.39\%$ less than that produced by NC. This is because during the graph construction phase we perform the minimum number of edge edits. As for Erdos, which is larger than the football network, our error is about $39\%$ less than that of NC. Finally, we used the large network Enron to test the scalability of our algorithm. The corresponding average error produced is $33.05\%$ less than that of NC. 

\section{Security Analysis}

The essential features of our generalized multi-objective approach are the flexibility and practicality while applying the least modifications needed for a graph to be anonymous. 
%It is possible to achieve anonymity even if we cannot find $k-1$ vertices other than $v$ whose degrees are exactly $degree(v)$ as in the more restrictive classical $k$-anonymity algorithms. 
Depending on the sensitivity of the data to be anonymized, the value of the parameter $t$ differs as it can be set to balance between privacy and utility. If we want to anonymize very sensitive data, then setting $t$ to zero and applying our anonymization approach gives better results in terms of preserving data utility than using traditional approaches. For each vertex, we are applying the minimum number of edge editing operations. However, if the data is not critical enough but the statistical information (for data mining purposes) is essential, we can assign a value for $t$ that keeps the data utility as we can see in Table \ref{tab:GrQc-results} where we present experimental results on the "GrQc" network when $t=0$ versus those that are found in \cite{univariate}. We measured the average error induced by each algorithm on the data utility measures for different $k$ values. When, $t=0$, parameterized kDA produced an average error that is 47 $\%$ less than that produced by NC on the harmonic mean of the shortest path. For transitivity, our average error is 10 $\%$ less than that of NC. Having this flexibility makes our approach a general umbrella for the existing algorithms because we can preserve data utility and fine tune by setting the parameter $t$.
%are preserving data utility by setting the parameter $t$. 
Of course, our approach focused on enhancing the preservation of data utility without affecting privacy.

\begin{table}[htb!]
\caption{Experimental results on the GrQc dataset using NC and our algorithm when $t=0$}
\resizebox{13 cm}{!}{
\begin{tabular}{llcll}
                      &                                                                                              & \multicolumn{1}{l}{}                    &                                     &  \\ \cline{2-4}
\multicolumn{1}{l|}{} & \multicolumn{1}{l|}{\textbf{Measure}}                                                        & \multicolumn{1}{c|}{\textbf{Algorithm}} & \multicolumn{1}{l|}{\textbf{Error}} &  \\ \cline{2-4}
\multicolumn{1}{l|}{} & \multicolumn{1}{l|}{\multirow{2}{*}{The   largest eigenvalue of the adjacency matrix}}       & \multicolumn{1}{c|}{NC}                 & \multicolumn{1}{l|}{0.134}          &  \\ \cline{3-4}
\multicolumn{1}{l|}{} & \multicolumn{1}{l|}{}                                                                        & \multicolumn{1}{c|}{Parameterized kDA}  & \multicolumn{1}{l|}{0.127}          &  \\ \cline{2-4}
\multicolumn{1}{l|}{} & \multicolumn{1}{l|}{\multirow{2}{*}{The second smallest eigenvalue of the Laplacian matrix}} & \multicolumn{1}{c|}{NC}                 & \multicolumn{1}{l|}{0.242}          &  \\ \cline{3-4}
\multicolumn{1}{l|}{} & \multicolumn{1}{l|}{}                                                                        & \multicolumn{1}{c|}{Parameterized kDA}  & \multicolumn{1}{l|}{0.223}          &  \\ \cline{2-4}
\multicolumn{1}{l|}{} & \multicolumn{1}{l|}{\multirow{2}{*}{Average Path Length}}                                    & \multicolumn{1}{c|}{NC}                 & \multicolumn{1}{l|}{0.097}          &  \\ \cline{3-4}
\multicolumn{1}{l|}{} & \multicolumn{1}{l|}{}                                                                        & \multicolumn{1}{c|}{Parameterized kDA}  & \multicolumn{1}{l|}{0.09}           &  \\ \cline{2-4}
\multicolumn{1}{l|}{} & \multicolumn{1}{l|}{\multirow{2}{*}{Harmonic mean of the shortest distance}}                 & \multicolumn{1}{c|}{NC}                 & \multicolumn{1}{l|}{0.15}           &  \\ \cline{3-4}
\multicolumn{1}{l|}{} & \multicolumn{1}{l|}{}                                                                        & \multicolumn{1}{c|}{Parameterized kDA}  & \multicolumn{1}{l|}{0.08}           &  \\ \cline{2-4}
\multicolumn{1}{l|}{} & \multicolumn{1}{l|}{\multirow{2}{*}{Transitivity}}                                           & \multicolumn{1}{c|}{NC}                 & \multicolumn{1}{l|}{0.03}           &  \\ \cline{3-4}
\multicolumn{1}{l|}{} & \multicolumn{1}{l|}{}                                                                        & \multicolumn{1}{c|}{Parameterized kDA}  & \multicolumn{1}{l|}{0.027}          &  \\ \cline{2-4}
\multicolumn{1}{l|}{} & \multicolumn{1}{l|}{\multirow{2}{*}{Sub-graph   centrality}}                        & \multicolumn{1}{c|}{NC}                 & \multicolumn{1}{l|}{0.757}          &  \\ \cline{3-4}
\multicolumn{1}{l|}{} & \multicolumn{1}{l|}{}                                                                        & \multicolumn{1}{c|}{Parameterized kDA}  & \multicolumn{1}{l|}{0.754}          &  \\ \cline{2-4}
                      &                                                                                              & \multicolumn{1}{l}{}                    &                                     &  \\
                      &                                                                                              & \multicolumn{1}{l}{}                    &                                     & 
\end{tabular}
}
\label{tab:GrQc-results}
\end{table}

To evaluate the robustness of our model in preserving privacy and security, we analyze and contradict the behavior of an intruder. If we consider the example of a medical data set where an intruder may have some information about a targeted individual like age, gender, or address, among other things collected from external resources. Then, the (potential) intruder tries to match this information with nodes in the released network. Using our constrained anonymization model to mask the data before releasing it has a better chance of hindering the intruder from identifying his/her target because the given information will be matched with many more individuals. To illustrate, if an individual in a network has 500 links, and $t=3$, then our approach can yield more vertices whose 
%there are at least $k-1$ 
degrees fall in the interval $[497,503]$. In practice, distinguishing between an individual with 500 links and another with 503 links is challenging. 
Furthermore, an intruder may try to link his/her a priori information with released nodes and estimate the probability of correct matching. In this context, a node is at risk if its identification probability exceeds a certain value. This probabilistic technique is particularly not helpful for identifying individuals in our case since we guarantee that many nodes have same identification probability, which obviously increases the intruder's uncertainty.

\section{Conclusion and Future Work}

We presented a generalized degree anonymization problem by adding constraints and bounds on the number of edge modifications, thereby relaxing the problem definition, which can extend its practicality and use in real applications. 

When applying degree-based anonymization we have to use graph realization to obtain the resulting graph. We interpreted the graph realization problem as a new variant of the Weighted Edge Cover problem and formulated it as an Integer Linear Programming problem. This has enabled us to compute the best possible solution by minimizing data loss caused by edge modification operations. The weighted variant of Edge Cover could be of interest on its own.

We considered simple unweighted undirected graphs, but this work can be extended to other types of graphs. Future work includes testing our approach on dynamic graphs, like online social networks where degrees vary with time, which requires adjusting the degree-anonymization solution. A possible approach would be to use a parameterized dynamic variant of the problem (see \cite{abu20,abu15, KrithikaST18,rolf-dynamic-cluster-editing} for recently studied parameterized dynamic problems). It would also be interesting to modify our notion of degree-based anonymity to apply to distributed networks like intranets or communication networks.

%There is no consensus on the definition of data utility or privacy of networks. Various metrics have been used to test data utility, so it is difficult to compare different algorithms in the literature. There is a critical need for privacy quantification models and metrics. 

%\bibliography{References}
%\bibliographystyle{abbrv}

\end{document}